\documentclass[11pt,twoside]{article}
\usepackage{ap-article}
\usepackage[utf8]{inputenc}
\usepackage[numbers]{natbib} 
\usepackage[english]{babel} 
\usepackage{amsmath}
\usepackage{amssymb}
\usepackage{amsfonts}
\usepackage{amssymb}
\usepackage{calligra}
\usepackage{calrsfs}

\usepackage{relsize}
\usepackage[mathscr]{euscript}
\usepackage{graphicx}
\usepackage{blindtext} 
\usepackage{epstopdf}
\usepackage{booktabs}
\graphicspath{ {./Figures/} }
\usepackage{listings}
\shortauthors{A. Alshehri et al.}
\shorttitle{Multi-Agent Communication under Epistemic Planning}
\title{Modeling Communication of Collaborative Multi-Agent System under Epistemic Planning}

\author{%
Abeer Alshehri\,\up{1,2}\,%\footnote{Typeset names in 10 pt Times Roman, boldface. Use the footnote to indicate the present or permanent address of the author.}\,,\
Tim Miller\,\up{2}\,
Liz Sonenberg\, \up{2}\
}
\addresses{%
\up{1}
    Department of Computer Science and Information Systems,\\
     King Khalid University, Abha, Saudi Arabia
\\ %\vspace*{0.04truein}
E-mail: adshehre@kku.edu.sa
\\ \vspace*{0.05truein}
\up{2}
 School of Computing and Information Systems,\\
     University of Melbourne, Victoria, Australia
     \\
E-mail: \{tmiller, l.sonenberg\}@unimelb.edu.au
}
\begin{document}\maketitle

\abstracts{%
In most multi-agent applications, communication is essential among agents to coordinate their actions and achieve their goals. However, communication often has a related cost that affects overall system performance. In this paper, we draw inspiration from epistemic planning studies to develop a communication model for agents that allows them to cooperate and make communication decisions effectively within a planning task. The proposed model treats a communication process as an action that modifies the epistemic state of the team. We evaluate whether agents can cooperate effectively and achieve higher performance using communication protocol modeled in our epistemic planning framework in two simulated tasks. Based on an empirical study conducted using search and rescue tasks with different scenarios, our results show that the proposed model improved team performance across all scenarios than baseline models. 
}

% Please include a maximum of seven keywords

\medskip
%-------------------------- KEYWORDS ---------------------------------------


\vspace*{7pt}\textlineskip
%-------------------------- BEGIN BODY OF TEXT -----------------------------
%\begin{multicols}{2}

\section{Introduction}

In many applications, multiple autonomous agents working as a team have potential advantages relating to the effective execution of tasks beyond the capabilities of a single agent. With interdependent tasks, effective collaboration and coordination are the key foundation of a higher-performing system. An agent has to be transparent to other agents to be able to coordinate their actions \cite{miller2018social}. 

One of the main requirements of effective collaboration is the existence of a communication process. In critical domains, such as disaster response and healthcare, the communication process among team members is considered an essential requirement for better team cooperation. In a multi-agent system (MAS), agents communicate to develop and maintain a shared mental model (SMM) on a task. An SMM is an internal representation of an environment that enables agents to form accurate explanations and expectations regarding the task and coordinate their actions to meet the demands of that task \cite{rouse1986looking}. Although communication is a natural way to increase the SMM of a team, excessive communication, which may arise due to computational requirements, has been costly and hinders system performance \cite{unhelkar2016contact}. Besides, the information obtained may not be valuable for the agent in making their decision. When faced with collaborative tasks, effective coordination through communicating at the right time with the right information becomes crucial to achieve a higher system performance.

Recently, studies have discussed different methods for improving multi-agent systems' performance, namely, the ability to reason regarding the exchanged information, and when and what the agent should communicate. By gaining insight into how humans cooperate effectively, research shows that an intelligent human interaction is a result of their ability to reasoning about others' mental states ---beliefs and knowledge--- and thus \emph{selectively} communicating based on the needs of team members \cite{lim2006team,mathieu2000influence,verbrugge2009logic}.

Most works in developing communication considered it to be in a separate model from the planning process; thus, it assumes teams work simultaneously to make planning and communication decisions at each step \cite{best2018planning,unhelkar2016contact}.  Since planning does not typically consider communication as an involved action, we believe that using an \emph{epistemic planning} framework leads to a planning process that can consider communication action.

There have been recent works on planning using \emph{epistemic logic} \cite{bolander2011epistemic,sonenberg2016social,engesser2017cooperative,schwarzentruber2019epistemic}. Since it is concerned with representing knowledge/beliefs rather than simple facts about an environment, it forms a strong basis from which to reason in the absence of complete knowledge . By enabling agents to reason in such a way, they have a compact class of planning that models communication as a natural action, allowing a planner to fold in communication planning during task planning \cite{miller2018social}. These actions are carried out if newly observed information is inconsistent with the agents' current beliefs and useful for other agents in achieving their tasks. This model could ultimately maintain better performance by enabling agents to communicate and keep track of SMMs in one compact model.

In this paper, we address the following research question: \emph{Does reasoning about knowledge of others (epistemic reasoning) in the presence of communication as an action improve the cooperation of a multi-agent system performing in a dynamic environment?} In simulated tasks motivated by disaster response and search and rescue, we evaluate whether agents can cooperate effectively and achieve higher performance using communication protocol modeled in our epistemic planning framework. 

The paper is structured as follows. Section 2 provides the relevant background and discusses related work.  We set out the proposed model in detail in Section 3. In section 4, we present the experiments that tested our approach. Then, section 5 discusses the results of the conducted experiments. Finally, our paper concludes with our work's contribution, along with some future directions of research.
%%%%%%%%%%%%%%%%%%

\section{Background}

In this section, we overview relevant prior work on multi-agent systems and epistemic planning framework related to effective team performance.

\subsection{Shared Mental Models (SMM)}

In the context of teamwork, the psychology literature has introduced the concept of a shared mental model (SMM) to explain how a human team works. It represents the knowledge of team members for the world around them that helps to understand and predict others' behavior and, in turn, adapt their actions to task demands \cite{rouse1986looking}. For instance, a blind pass between members of a basketball team shows that each member can predict others' position on the court. Extensive teamwork experiments have shown that higher similarity of team mental models results in higher team performance \cite{harbers2012enhancing,harbers2012measuring,mohammed2010metaphor}.

The ability to predict others' behavior allows anticipating actions to their expected behavior along with team tasks. Team implicit coordination relies on having common knowledge that assists in predicting team behavior and needs, including beliefs and goals along with a sharing understanding of the team task \cite{converse1993shared,entin1999adaptive}.

Maintaining SMMs developed in the psychological literature can be applied to human-agent or agent-agent systems to improve their work. Having a shared mental model about the team functioning will improve its performance \cite{lim2006team,mathieu2000influence}. Although the word ``shared'' suggests identical mental models, this is not the case where autonomous agents are distributed in a dynamic environment. Instead, studies have proposed similarity measures to investigate the degree of similarity for good team performance or good enough performance to achieve specific tasks  \cite{bolstad1999shared,lim2006team,mathieu2000influence,harbers2012measuring,singh2016communication}. In this work, maintaining an SMM enables agents to share information efficiently and achieve a better team performance.

\subsection{Communication Strategies} 

The essential process for maintaining and developing SMMs is the communication between team members. However, excessive exchanging of information results in a communication overhead, which will negatively affect team performance. Due to an agent's limited view of an environment, it might be unwanted or impractical to share all knowledge all the time. Communication has a related cost associated with the required resources or bandwidth. Thus, effective team performance results from evaluating the trade-off between communication cost and the value of information received. That leads to more efficient communication by which a common ground (i.e., SMM) is developed among team members to better coordinating implicitly \cite{miller2017logics}. 
%Studies have shown that teammates who communicate based on anticipating others’ needs have a higher performance than those who have less anticipation \cite{entin1999adaptive,macmillan2004communication}. Additionally, \citeauthor{macmillan2004communication} \cite{macmillan2004communication} proposed that the nature of the task and its environment determine the implicit and explicit coordination among team members. It has also been found that there is a relationship between the workload and the types of communication, a higher workload leading to less exchange of information among teammates \cite{kleinman1989team}.

There are some factors for effective communication---what information to communicate, when, and to whom. \citeauthor{kinney1998learning} \cite{kinney1998learning} argued that there is a condition where the environment may affect the agents' mental states if they work using a static communication protocol. Handling the observability and environmental changes is required for cooperative agents. They also argued how an agent evaluates information before sharing it to determine its usefulness for both the task and other agents in order to avoid the cost of consuming communication resources. In addition, \citeauthor{li2015communication} \cite{li2015communication} and \citeauthor{harbers2012enhancing} \cite{harbers2012enhancing} observed that communicating goals (intentions) contributes more to team efficiency in human-agent teams than sharing beliefs (world knowledge) to build a strong SMM. Regarding certain classes of knowledge that need to be communicated, it is better to consider what beliefs should be shared and whether they are sufficient for the team. 

Other work has concentrated on the relationship between task structure, anticipatory information sharing, and team performance. \citeauthor{butchibabu2016implicit} \cite{butchibabu2016implicit} proposed a set of hypotheses to identify these relationships. They evaluated them through empirical study, including multiple teams of four people in a collaborative search-and-deliver task within a simulation environment. They analyzed the strategies for effective team communication in tasks with different complexity levels. They observed higher rates of implicit coordination among the best performing team rather than explicit coordination. This was true even with a high-complexity task as a result of training teams before starting experiments. Furthermore, they identified that proactive communication between team members regarding the next goal reduces communication costs and improves team performance. However, these results may not cover the case where teams have not worked together before or lack understanding of the task.

\citeauthor{unhelkar2016contact} \cite{unhelkar2016contact} developed a communication policy, ConTaCT, which enables agents to decide whether to share information with others or not during time-critical collaborative tasks within a simulated task motivated by rescue operations. They evaluated its performance within a deterministic domain (partially observable states) with a partially-known initial state. The algorithm enables an agent to maintain other agents' knowledge about the environment and then compare the benefits of this decision against communication cost. They found that ConTaCT reduces communication overhead within artificial agent teams while achieving a comparable task performance, reducing more than 60\%. The reason is that some information is unnecessary for other members and benefits only the local agent. 

\citeauthor{zhou2012communication} \cite{zhou2012communication} worked on communication problems related to the question of why and when to communicate. They introduced communication to the interactive, partially observable Markov decision processes (I-POMDPs). They presented a communicative, interactive POMDPs (Com-I-POMDPs) by integrated costly communication actions into (I-POMDPs) framework. Their results show a graphical model of the framework that consumes less time and space to solve a problem in uncertain multi-agent settings. More recently, \citeauthor{gmytrasiewicz2019optimal} \cite{gmytrasiewicz2019optimal} built on the I-POMDP framework and included the communication process to formulate Communicative I-POMDPs (C-IPOMDPs). Similar to our work, communication is treated as an action. They use the Bellman optimality principle as decision-theoretic planning to deciding whether they are communicative acts, just as the other actions. However, their framework has not been tested to verify their findings.

%--------------------------------------------------------------------------------------

\subsection{Evaluation of Team Performance and Communication} 

Several studies have shown a critical need for a solid understanding of what influences team interactions and performance \cite{hall1996cognitive,entin2000assessing}. Here we describe standard methods that are used to evaluate team performance and communication. Analyzing these methods would help in identifying ways to improve team performance and communication.

Methods of evaluating team performance used in prior research have been tailored to specific scenarios and involved objective measures such as a total number of accomplished tasks and accuracy rates, or subjective measures such as task uncertainty and the workload assigned to each agent \cite{goldman2003optimizing,harbers2012enhancing,unhelkar2016contact,entin2000assessing,singh2016communication}. In this paper, team performance is measured based on the total number of actions it takes to complete the task (completion cost) and task completion time.

Measuring communication behavior in prior studies was based on the total number of communications per minute or per task \cite{entin1999adaptive,harbers2012enhancing,unhelkar2016contact,singh2016communication}. \citeauthor{macmillan2004communication} \cite{macmillan2004communication} measured communication efficiency based on anticipation ratio, which is the ratio of total communications to communication requests. The higher the team shows the anticipation ratio, the more efficient communication, and implicit coordination. We measured communication behaviors based on the total number of messages exchanged by the agents over a task in our work.

SMMs have been used to predict and understand team performance. Agents communicate as soon as they obtain new observations in order to develop an SMM for a task. Measuring the degree (level) of sharedness of mental models that ensure good team performance is important, that is, which beliefs of mental models should be shared?  \citeauthor{harbers2012measuring,jonker2011shared} \cite{harbers2012measuring,jonker2011shared} proposed similarity measures that can be applied to intelligent agents. \citeauthor{jonker2011shared} \cite{jonker2011shared} defined what is called (subject overlap) of agents' mental models to determine the extent of their similarity. It is a measure of the percentage of answers agents were provided to a set of queries related to a task. Based on this, we measured the sharedness of mental model contents (i.e., agents' beliefs) over a task.

%----------------------------------------------------------------------------------------

\subsection{Dynamic Epistemic Logic (DEL)}

Epistemic logic (EL) is a means of handling knowledge, a concept that was first published by \citeauthor{melb.b126434919620101} \cite{melb.b126434919620101}. Having knowledge something is true (false) means either being true (false) in every state that is considered to be possible or being uncertain about it (i.e., true in some states and false in others). A key merit of EL is that the state of knowledge for several agents can be represented using a Kripke structure \cite{fagin1995my}. Given a set of propositional variables, Kripke structure forms a set of possible states (epistemic alternatives), a set of accessibility relations (binary relations on the world states), and a set of propositional facts (possibilities for the truth of propositional variables in each world state) \cite{van2007epistemic}. 

%For instance, if there are two agents, a and b, who know nothing about a specific aspect of world states, whether p and q hold, a Kripke structure will present the possibilities for truth to be ($\oldemptyset$, p, q, pq). The accessibility relation of agent a is $\sim$a in these four possibilities, and $\sim$b for agent b. Such a model allows us to represent and reason with simple knowledge as well as with the higher-order knowledge of all relevant agents concurrently, e.g. agent b knows that agent a is unsure whether q is true \cite{jensen2014epistemic}.

Extending epistemic logic with dynamic aspects has been studied over recent decades under the term dynamic epistemic logic (DEL) \cite{fagin1995my,baltag2016logic,costa2002first,li2009information}. In DEL, changing knowledge is modeled by transforming Kripke structures. These transformations usually involve factual (or ontic) changes and changes in the agents’ mental states. This means that agents’ accessibility relations have to change, and accordingly, the set of possible states may change as well \cite{van2007epistemic}.

%----------------------------------------------------------------------------------------
\subsection{The DEL Planning Framework}

To achieve a specific goal in a particular situation, we need to organize our actions in a plan which represents how we should act. In a planning task, otherwise known as a planning problem, three components are given: an initial state, a set of available actions, and a set of goal states. A planning solution is a sequence of actions that converts the initial state into one of the goal states \cite{majercik2003contingent}.

A method for generating plans works by predicting the outcome of actions that lead to the goal. There are many approaches to planning systems, mostly connected to propositional logic \cite{geffner2013concise}. Epistemic planning is one such approach, which uses epistemic logic as the underlying formalism \cite{bolander2011epistemic,lowe2011planning,andersen2012conditional} - we refer to \cite{bolander2017gentle} for an overview of DEL-based epistemic planning. Generally, epistemic planning considers how to get from a current state of knowledge to a goal state. The three components of the DEL-based planning approach are an epistemic model as the initial state, a set of event models (or action models) to describe the actions, and an epistemic formula to describe a goal. The solution for such a planning problem is a sequence of event models that uses product update operators to produce a model that meets the goal formula \cite{bolander2011epistemic,baltag2016logic}.

Planning in partially observable environments is computationally challenging. Partially observable Markov decision processes (POMDP) are a common approach to planning in such an environment \cite{kaelbling1998planning,melo2011pomdp,amato2015probabilistic}. However, the POMDP approach has limited applicability to planning under incomplete knowledge compared to epistemic logical approaches \cite{herzig2003action}. The latter approaches allow for more compact modeling of planning tasks about \emph{nested knowledge} than POMDPs; that is, knowledge about what other agents know. In the DEL-based planning approach, actions (event models or action models) are considered a feature that encodes partial observability and non-determinism \cite{andersen2012conditional}. Thus, observability will be based on executing actions, and there is no need to specify an observation function as an action description.

Furthermore, the DEL-based planning approach allows us to formalize reasoning about knowledge or belief changes (including higher-order knowledge or beliefs about other agents). Such reasoning is required for many AI applications that involve multiple autonomous agents interacting to accomplish collaborative or competitive tasks.

%----------------------------------------------------------------------------------------
\subsection{Epistemic Planning for a Cooperative Multi-agent System
}

In multi-agent situations, epistemic planning can be used for decision making in the case of distributed knowledge. It is essential for successful collaboration and coordination that agents can reason about other agents' knowledge, capabilities, and uncertainty. Such a planning framework integrates non-determinism and partial observability in multi-agent tasks \cite{aucher2013undecidability}.

Multi-agent planning tasks can be modeled using Kripke structures as world states and event models as actions. We have to consider some additional design choices for such tasks, such as whether agents are collaborating or competing; or communicating arbitrarily and committing to a joint plan. In a decentralized setting, \citeauthor{bolander2016better,engesser2017cooperative} \cite{bolander2016better,engesser2017cooperative} proposed implicitly coordinated plans that help to solve tasks with joint goals where agents at the time of planning are not allowed to commit to a joint policy. \citeauthor{muise2015planning,kominis2015beliefs} \cite{muise2015planning,kominis2015beliefs} modeled epistemic planning problems with bounded nesting of knowledge (or belief) and compiled this into a form that classical planning can handle. Such a model can plan from a single agent's viewpoint with the possibility for goals and actions having nested beliefs, co-present observations, and taking other agents' beliefs into account while reasoning. In our work, we evaluate the efficiency of using an epistemic planning framework to model the communication process using an epistemic planner proposed by \citeauthor{muise2015planning} \cite{muise2015planning}.
%%%%%%%%%%%%%%%%%%%%%%%%%%%%%%

\section{Method}
In this section, we describe the objective of this work. We describe an epistemic planning task of multi-agent settings. Then we lay out the proposed model (Selective communication Model) in detail.

\subsection{Objective}

Our model's objective is to incorporate a cooperative team with varying tasks, in which the planning task is categorized under the epistemic planning framework. This will enable an agent to model their beliefs and other beliefs; thus, aligning SMM to keep track of consistent beliefs. Based on this, agents will communicate \emph{selectively} only in case of inconsistent beliefs that are newly observed from a current state and valuable for other members to achieve their tasks. We aim to investigate the impact of applying such a framework on team performance and compare it with two generated baseline strategies.

\textit{Hypothesis:} We hypothesize that a team can communicate more \emph{selectively} with minimal loss of performance if they represent its SMM compared with baseline communication models. Having a model of others' beliefs will help them to collaborate effectively.

\subsection{Multi-agent Planning Task}

A model consists of $\mathscr{P}$ a finite set of epistemic literals, $\mathscr{A}{g}$ a finite set of cooperative agents, and $\mathscr{A}$ a finite set of actions. We can define a multi-agent epistemic planning (MEP) task $\mathscr{D}$ as follows:\\
$\mathscr{D}$ is a tuple of the form ($\mathscr{P}$, $\mathscr{A}$, Ag,  $\mathscr{G}$), where $\mathscr{P}$, Ag and $\mathscr{A}$ are as above, and $\mathscr{G}$ is the goal condition. Each action $a \in$ $\mathscr{A}$ is assumed to be of the form $\normalfont\calligra {precondition}(a)$, $\normalfont\calligra {effect}(a)$, where both are finite sets over $\mathscr{P}$. An action a is said to be applicable in a state if $\normalfont\calligra {precondition}$ $\textsuperscript{+} (a) \subseteq$ $\mathscr{P}$ and $\normalfont\calligra {effect}$ $\textsuperscript{-} (a) \cap$ $\theta$. 
The set of formulae, $\mathscr{L}$, for multi-agent
epistemic logic is given by the following grammar \cite{muise2015planning}:

\begin{center} $\phi$ ::= \textit{p} $\vert$ $\phi$ $\wedge$ $\phi$ $\vert$ $\neg$$\phi$ $\vert$ $B_{\textit{i}} \phi$ $\vert$ $[\alpha] \phi$ \end{center}

where \textit{p} $\in$ $\mathscr{P}$, \textit{i} $\in$ $\mathscr{A}{g}$, $B_{\textit{i}} \phi$ is that agent \textit{i} believes $\phi$, and $[\alpha] \phi$ is believed $\phi$ after applying action $\alpha$. The epistemic planning process has been approached by compilation to classical planning that deals with bounded
nesting of knowledge/belief. For more details, we refer to \cite{muise2015planning}.

We designed two tasks for Gridworld and Blocks World for Teams (BW4T) maps to test the effect of modeling communication as a natural action under an epistemic planning framework on team performance. A variation of task scenarios has been used to realize the validity of the proposed model in some different scenarios motivated by search and rescue operations. We assume that actions' effects are deterministic and commonly known for these tasks, but the environment is often initially unknown. Common knowledge of the planning behavior, initial state, and goal state is assumed to be known by the collaborator agents.

%%%%%%%%%%%

\subsubsection{Selective Communication Model}
 
Communication process is modelled as an action $a = \{precondition(a), effect(a)\}$ $\in$ $\mathscr{A}$, in which the $precondition(a)$ and $effect(a)$ are with respect to belief. The Selective communication model for a cooperative system allows each agent to communicate during a task as a part of the planning process. The following example will describe the model in details:

In a mission of searching for survivors on a map in the Gridworld domain (refer to section 4.1.1 for more details), agents are moving around to find the survivor locations and update the team's mental model about it. The decision of updating others' mental state involves the following three parts: (1) \emph{what to communicate}; (2) \emph{to whom should be communicated}; and (3) \emph{when to communicate}. Communication action $a \in$ $\mathscr{A}$ is modeled to answer them as the following:

\begin{itemize}
    \item $precondition(a)$ = $B_{\textit{i}} s_{1}$, an agent $i$ believes that there is a survivor $s$ at location 1. 
    
    \item $effect(a)$ = \{$\forall$ $n \in \mathscr{A}{g}$, $B_{\textit{n}} s_{1}$\}, for all other agents they believes that there is a survivor $s$ at location 1.
    
\end{itemize}
In the action's precondition, we model (\emph{what to communicate}) part of the communication decision by determining a proposition that needs to be communicated - one communication action for each proposition; that is here, the survivor location on the map. In the action's effect, we model (\emph{to whom should be communicated}) part by selecting the receiver, either all agents or specific agents -based on the task scenario, as we will see later (refer to section 4). As the planner models the effect of actions in $\mathscr{A}{g}$, that allows performing epistemic reasoning. Besides, it is planning from the perspective of a single agent (root agent) to form a coordinated plan accomplished by multiple agents \cite{muise2015planning}. Therefore, deciding (\emph{when to communicate}) is the planner ability in which the information is communicated only just as needed to maintain consistent beliefs (SMM) essential for them to achieve their task.

\subsection*{Example} Here is the communication action $commsurvivor$, to communicate survivor location, coded in PDDL for Gridworld domain where the agents are required to search positions of the three survivors on the map. We use the types \verb!?p! for positions, \verb!?a! for agents and \verb!?s! for survivors. We introduce fluents \verb!(at ?a ?p)! to
represent that agent \verb!?a! is at position \verb!?p!, and \verb![?a](survivorat ?s ?p)! to represent that the agent \verb!?a!
believes that survivor \verb!?s! is at position \verb!?p!. The effect of that action is to change the belief state of other agents \verb!?g! to all believe that survivor \verb!?s! is at position \verb!?p! :

\begin{lstlisting}
(:action commsurvivor
  :derive-condition  always
  :parameters        (?p - pos ?a - agent  ?s 
                      - survivor)
  :precondition      (and (at ?a ?p) [?a]
                        (survivorat ?s ?p)) 
  :effect            (and (forall ?g - agent 
                        [?g](survivorat ?s ?p
                     )))
)
; ...
\end{lstlisting}

\subsubsection{Constructing a Shared Mental Model (SMM)}

A mental model of a particular agent, perhaps a root agent who perceives the environment, represents a state of the world; The planner's reasoning is from the perspective of that agent, what its true and false beliefs \cite{muise2015planning}. Action execution, then, is grounded on these hold beliefs (action's precondition). A Shared Mental Model (SMM) construction represents the overlapping content of agents' mental models; that is, how many propositions are always known among agents. It is updated based on the actions' effects. Thus, communication action's effect will inform others about what has been observed and the communicator agent's current location on the map. Additionally, the SMM can be updated if more than one agent observes the same state; that is, they all have the same beliefs (a common knowledge) about that state.

 %Consequently, an individual agent's communication decision is a result of reasoning SMM, in which the agent can decide what to do next.

%When agents construct the SMM over a task, even though there is no shared information, they could estimate each others' mental model and \emph{selectively} communicate what is essential for them to achieve their sub-task.
%%%%%%%%%%%%%

\section{Evaluation}

In this section, we first explain an experimental design for the two simulated tasks used to empirically evaluate our proposed model. Then we present a set of measurements proposed to measure team performance.

\subsection{Experimental Setup}
We conducted a study involving two team compositions: a 3-agent team and a 4-agent team. Each team performed two tasks with five different scenarios (S1, S2, S3, S4, S5): epistemic goal, non-epistemic goal, commander (broadcast communication), commander (non-broadcast communication), and blocked cells. These tasks run within two simulated domains: Gridworld and Blocks World for Teams (BW4T). We used two map sizes for Gridworld domain: 3×3 and 4×3; two map sizes for Blocks World for Teams (BW4T) domain: three rooms and six rooms. The setups are as shown in Tables \ref{tbl:3} and \ref{tbl:4}.

We employed three communication models: (1) Selective communication model where agents can communicate based on SMM reasoning; (2) No Communication baseline model where agents cannot communicate as we dropped that action from the domain file; and (3) Communication All baseline model where agents must communicate every new observation --that is, they do not track an SMM.

We ran five task scenarios for each communication model for two maps combined with two team sizes, giving us 60 combinations. All the experiments were performed in the epistemic planner proposed by \citeauthor{muise2015planning} \cite{muise2015planning}. These instances were run on Intel Core i7-4770 CPU @ 3.40GHz.

\begin{table*}
\centering
   \caption{Experimental setups for Gridworld domain}   \label{tbl:3}
\small
\begin{tabular}{lcccccc}  
\toprule 
\textbf{Communication model} & \multicolumn{2}{c}{Selective} & \multicolumn{2}{c}{No Communication} & \multicolumn{2}{c}{Communication All}\\
\midrule
\textbf{Map size} &   3×3 & 4×3 & 3×3& 4×3 & 3×3 & 4×3 \\ 
\textbf{Team size} &  3 agents & 4 agents & 3 agents & 4 agents & 3 agents & 4 agents \\
\textbf{Task scenarios} &    S1 .. S5 & S1 .. S5 & S1 .. S5 & S1 .. S5 & S1 .. S5 & S1 .. S5\\
\bottomrule
\end{tabular}
\end{table*}

\begin{table*}
\centering
   \caption{Experimental setups for Blocks World for Teams BW4T domain}   \label{tbl:4}
\small
\begin{tabular}{lcccccc}  
\toprule 
\textbf{Communication model} & \multicolumn{2}{c}{Selective} & \multicolumn{2}{c}{No Communication} & \multicolumn{2}{c}{Communication All}\\
\midrule
\textbf{Map size} &   3 & 6 & 3 & 6 & 3 & 6 \\ 
\textbf{Team size} &  3 agents & 4 agents & 3 agents & 4 agents & 3 agents & 4 agents \\
\textbf{Task scenarios} &    S1 .. S5 & S1 .. S5 & S1 .. S5 & S1 .. S5 & S1 .. S5 & S1 .. S5\\
\bottomrule
\end{tabular}
\end{table*}

To complete a task successfully, agents need to plan both their actions and communications. A sequential plan for such a task is a sequence of actions where executing them will lead to a goal. The planner is responsible for allocating tasks for agents. For simplicity, we use a sequential planner and assume that every agent takes one turn after another.

\begin{Figure}
\includegraphics[width=3cm]{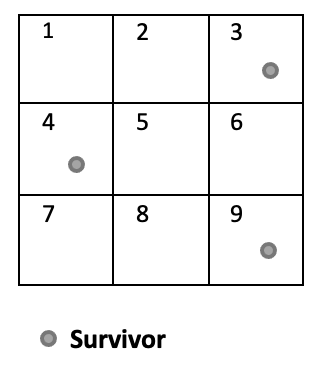}
 \centering
  \fcaption{The Gridworld (3×3) map.}     \label{fig:2}
 \centering
\end{Figure}

\subsubsection{Gridworld Domain - Search Task}

Gridworld is a simulated world, with a predetermined size and obstacles, consisting of a rectangular subset of the integer plane \cite{boddy1989solving} (Figure \ref{fig:2}). Each point is a position that is occupied by no more than one agent at a time. The agent can move onto any of the free neighbors of the currently occupied position. The distance between adjacent positions is equal to one unit. Each agent has a map of the world and can find their way to the goal with a planned path. Planning an agent's path is beyond the scope of this paper. As our tasks are motivated by search and rescue operations, we modified the Gridworld domain by adding a specified number of survivors (three survivors) in different positions. Thus agents are required to search positions of them on the grid. Agents do not have prior knowledge of the distribution of survivors, and thus, they must observe the grid and communicate effectively. At the outset, agents are distributed at known starting positions. 

Each agent must perform a set of actions that we encoded within a domain file to accomplish the goal.  To move in any direction, agents used
the \emph{move} action. Additionally, they used the \emph{observe} action to observe a survivor's position and the \emph{communicate} action to broadcast it to others and update their beliefs.

\paragraph{Epistemic Goal Scenario}
In this scenario, we specified the goal in an epistemic form related to agents' beliefs. Therefore, the goal was to search the whole grid and make all agents know all survivors' positions.

\paragraph{Non-epistemic Goal Scenario}
In this scenario, we specified the goal in a non-epistemic form related to simple facts about an environment. Therefore, the goal was only to search for the whole grid.

\paragraph{Commander (broadcast communication) Scenario}
In this scenario, we specified one agent as a commander who always stays in one position and does not actively search. Therefore, the goal was to have agents check all positions and have only the commander be aware of all survivors' positions. This means that individual search agents will observe the survivors' positions, even if that is not a part of the goal, and then communicate them to the commander (broadcast communication).

\paragraph{Commander (non-broadcast communication) Scenario}
Like the previous scenario, the only difference was that only one individual search agent could communicate all survivors' positions to the commander (non-broadcast communication).

\paragraph{Blocked Cells  Scenario}
In this scenario, we added another dimension of uncertainty by specifying one survivor position and two cells to be blocked. Agents do not have prior knowledge about the distribution of both the survivor and blocked cells. Therefore, the goal was for agents to reach a specific position while searching for the survivor's position and avoiding blocked cells.

\subsubsection{Blocks World For Teams (BW4T) Domain - Search and Retrieval Task}

Blocks World For Teams (BW4T) is a simulated environment that consists of a number of rooms in which colored blocks are distributed randomly \cite{johnson2009joint} (Figure \ref{fig:3}). Agents must search, locate, and retrieve blocks from rooms and put them in a drop-zone in a specific order. An agent can only carry one block at one time. At the outset, agents are distributed at known starting locations. They operate within a potentially unknown environment. They know the locations of rooms (i.e., the map), but they do not know the blocks' location. This knowledge is obtained when the agent visits rooms that have these blocks and observe. Thus, agents need to communicate with each other to achieve this task effectively.

\begin{Figure}
  \includegraphics[width=4.5cm]{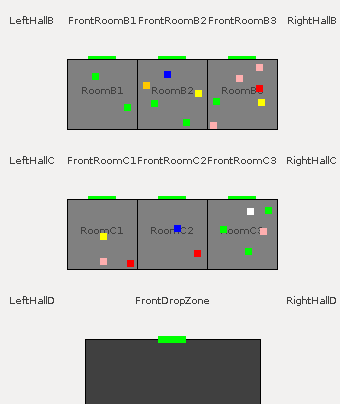}
  \fcaption{Blocks World for Teams BW4T (six rooms) map.}     \label{fig:3}
\end{Figure}

As our tasks are motivated by search and rescue operations, agents must search in rooms and retrieve two specific color blocks (higher priority elements to be rescued) and bring them to the drop-zone. There is a total of four blocks of various colors on the map. 
Each agent was required to perform a set of actions that we encoded within a domain file to accomplish the task, corresponding to the actions in the BW4T simulator. The agent uses the \emph{goTo} action and the \emph{goToDrop} action when they deliver a block to the drop-zone to move between rooms. To pick up the block, they use the \emph{pickUp} action and \emph{putDown} action to drop it at the drop-zone. They use the \emph{observe} action to observe and believe survivor's positions. To broadcast it for other agents and update their beliefs, they use the \emph{communicate} action.

\paragraph{Epistemic Goal Scenario}

In this scenario, we specified the goal in an epistemic form related to agents' beliefs. Therefore, the goal was to search rooms and make all believe that required color blocks are brought to the drop-zone.

\paragraph{Non-epistemic Goal Scenario}
In this scenario, we specified the goal in a non-epistemic form that relates to an environment. Therefore, the goal was to search in rooms and bring the required color blocks to the drop-zone.

\paragraph{Commander (broadcast communication) Scenario}
In this scenario, we specified one agent as a commander who always stays in one location and does not actively search. Therefore, the goal was for agents to search all rooms, and the commander believes that all required color blocks are brought to the drop-zone. This means that individual search agents observe the required color blocks' locations and bring them to the drop-zone, even if that is not a part of the goal. Then they must communicate it to the commander (broadcast communication).

\paragraph{Commander (non-broadcast communication) Scenario}
As in the previous scenario, the only difference was communicating with the commander as it has to be restricted to one known agent who allowed to share the commander what is being delivered (non-broadcast communication).

\paragraph{Blocked Cells Scenario}
In this scenario, we added another dimension of uncertainty by specifying three-color locations and one blocked room where agents do not have prior knowledge about the distribution of both blocks and blocked room. Therefore, the goal was for agents to search in rooms and bring two required color blocks to the drop-zone while observing a blocked room.

\subsubsection{Baseline1: No Communication Model}
 No Communication baseline model is implemented by eliminating explicit communication action from a domain file in some scenarios, including epistemic and non-epistemic goal scenarios, or by restricting communication action to a place where a commander stays, that is, for commander scenarios. This baseline model is motivated by some prior work \cite{wei2014role,unhelkar2016contact}.

In this baseline, despite eliminating explicit communication process, agents can still obtain some information about their team members due to interference among agents in their shared environment. Namely, where agents have common knowledge at the beginning of a task and a shared plan, they will not go through the same state twice.

\subsubsection{Baseline2: Communication All Model}
Communication All baseline model is implemented by forcing the planner to communicate each time an agent obtains a new observation. This would give us an upper bound for applying an epistemic framework to building an effective communication model. That is what would happen without having an SMM.

\subsection{Data Collection}

We collected data on various items of simulation experiments, measuring the following:
\begin{itemize}
\item Completion Time: Time it takes a team to solve a planning task.
\item Total number of actions: We measured the total number of actions it takes a team to complete a task.
\item Total number of communications: We measured the total number of messages exchanged by agents on a task.
\item Level of sharedness: We measured the consistency between agents' mental models over a task.
\end{itemize}

To illustrate how we measured SMM on a task using the concept of subject overlap (refer to Section 2.5), we used an example from the Gridworld domain (epistemic goal scenario). The map consists of positions and survivors; thus, we constructed some queries $\mathscr{Q}$ regarding, e.g., observed positions and observed survivors. For instance, does position 1 have a survivor? The answer to such queries will be compared with answers given by other models. Given that there are nine positions, for each position, whether it is observed or not, and whether it has a survivor or not, we formulated 36 queries related to the goal. Taking this set of queries $\mathscr{Q}$, we calculated the number of queries answered by all agents' mental models, which was three out of 36. The subject overlap was 3/36 = 8.33\% (as shown in Table \ref{tbl:1}).

In some cases, SMM is articulated into sub-models; thus, we also measured SMM between pairs, as in commander task scenarios.

We used a completion time and a total number of actions collectively as a proxy for team performance. The total number of communications indicates the coordination level among agents. We measured the level of sharedness for all runs to see its effect on team performance, particularly in the case of the Selective communication model. Also, we measure an average ratio as a relative effectiveness measure between approaches. 
%----------------------------------------------------------------------------------------

\section{Results}

In this section, we first report our findings from our simulation experiments that evaluate the proposed model's efficacy against two generated baselines. Then we discuss the results of the case studies in detail. We end with a discussion of our findings.

The team's performance was assessed based on a number of measurements (refer to Section 4.2). The overall performance was measured in terms of completion time and the total number of actions collectively. We collapsed the results of the five task scenarios for each domain for simplicity and gave the average completion time and the total number of actions. The low completion time and the total number of actions indicate high team performance on the task. Overall, teams using the Selective communication model exhibited, on average, lower completion times across all task scenarios with shorter plans (total number of actions) compared with the two other baselines (shown in Figures \ref{fig:5} and \ref{fig:6}). We show the relative effectiveness between models using a percentage change of the Selective-to-No Communication model and the Selective-to-Communication All model. A higher percentage indicates a lower team performance. In the Gridworld domain, as shown in Figure \ref{fig:5}, the trend percentages reflect worse performance for baseline models compared to the Selective model by 19\% and 119\% respectively in terms of completion time, and by 24\% and 3\% respectively in terms of the total number of actions. For the BW4T domain, as shown in Figure \ref{fig:6}, the trend percentages reflect worse performance for baseline models compared to the Selective model by  18\% and 12\% respectively in terms of completion time, and by 37\% and 8\% respectively in terms of the total number of actions.

\begin{Figure}
\begin{minipage}[b]{0.47\linewidth}
\centering
\includegraphics[width=\textwidth]{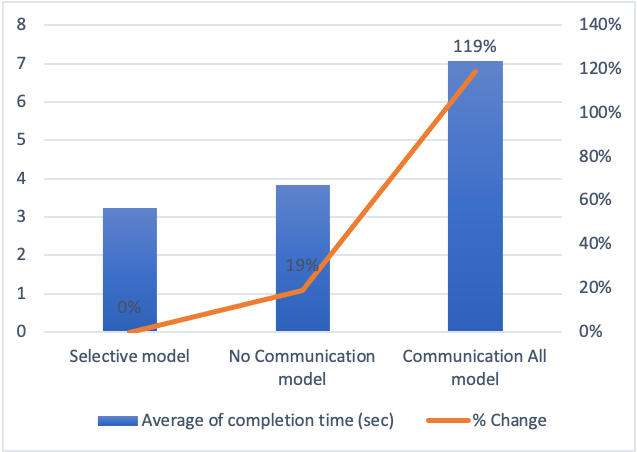}
\end{minipage}
\hspace{0.1cm}
\begin{minipage}[b]{0.47\linewidth}
\centering
\includegraphics[width=\textwidth]{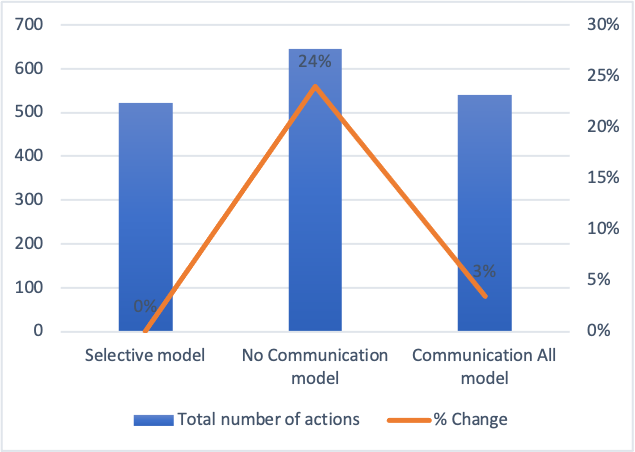}
\end{minipage}
\fcaption{The team performance in Gridworld domain, higher is worse}
\label{fig:5}
\end{Figure}

\begin{Figure}
\begin{minipage}[b]{0.47\linewidth}
\centering
\includegraphics[width=\textwidth]{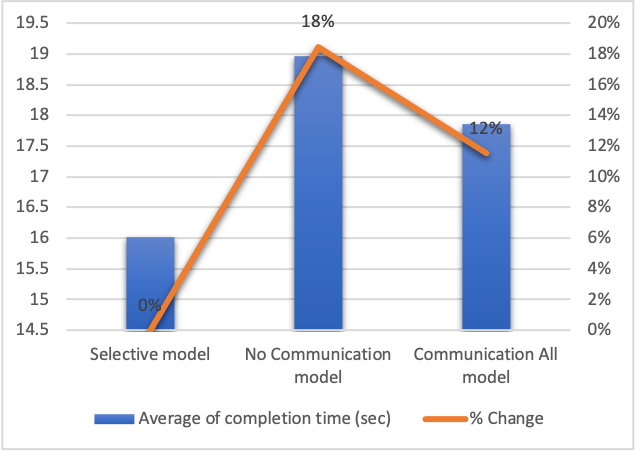}
\end{minipage}
\hspace{0.1cm}
\begin{minipage}[b]{0.47\linewidth}
\centering
\includegraphics[width=\textwidth]{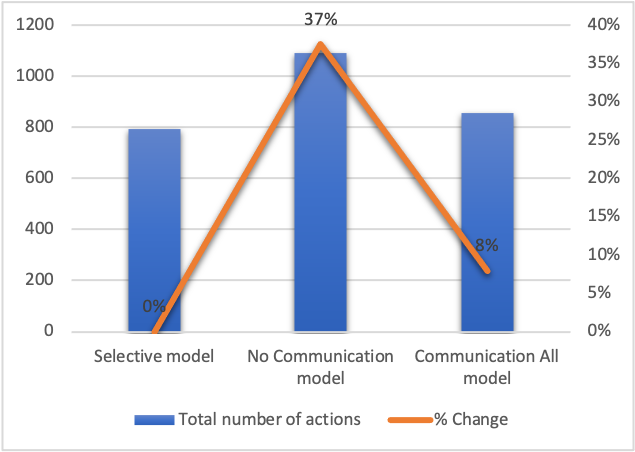}
\end{minipage}
\fcaption{The team performance of Blocks World for Teams (BW4T) domain, higher is worse}
\label{fig:6}
\end{Figure}

We hypothesized that a team could communicate more \emph{selectively} if the members represent an SMM with minimal loss of performance compared with baseline communication models. Generally, we found a significant effect of presenting the SMM on the Selective communication model's team performance. In the two domains, as shown in Tables \ref{tbl:1} and \ref{tbl:2}, presenting SMMs helps agents to communicate \emph{selectively} as the model shows better performance in terms of the total number of actions and communications compared with the two baseline models. That is also with a marginally better time completion of the tasks. However, the fewer messages presented by the No Communication model comes at the cost of the total number of actions. Agents using the No Communication model need to observe each state to get information, resulting in redundant actions. The Communication All model shows a comparable performance in terms of the total number of actions compared with the Selective model. However, that comes at the cost of a higher number of communications and completion times. Agents using the Communication All model communicate each new piece of information, resulting in unnecessary communications, exchanging information not required for other agents to complete their tasks.

The Selective model's performance comes from agents' ability to communicate when necessary to keep track of SMMs, thus, improving the team's overall performance. These findings are in line with \citeauthor{unhelkar2016contact} \cite{unhelkar2016contact}, who found that communicating only valuable information in cooperative tasks helps to maintain comparable performance with fewer communications. However, \cite{unhelkar2016contact} implemented the communication process separately from the planning model. Our proposed model utilized the epistemic planner \cite{muise2015planning} to model communication as an action; thus, the planning and communication occur in one compact model.

\subsection{Scenario Results} 	 		

Tables \ref{tbl:1} and \ref{tbl:2} show detailed results for each domain case study---Gridworld and BW4T---over five task scenarios (refer to Sections 4.1.1 and 4.1.2 for task details). 

\subsubsection{Epistemic Goal Scenario}
In this scenario, it can be seen from Tables \ref{tbl:1} and \ref{tbl:2} that the performance of the Selective communication model is marginally better in terms of completion time compared with the Communication All model, which has roughly the same number of actions and communications. In the No Communication model, it is clear that despite lower completion times compared to other models, it is more costly in terms of the total number of actions.

\subsubsection{Non-epistemic Goal Scenario}
In this scenario, Selective communication and No Communication models show a roughly similar performance. However, the latter is slightly better in completion time since there is no need to communicate as the goal is only to search. Agents do not need to tell each other the value of positions on the map to achieve the task, and if they do so, the team performance will worsen, as we see in the Communication All model.

\subsubsection{Commander (broadcast communication) Scenario}

This scenario shows that the Selective model's team performance is better than the baseline models. Agents using the No Communication model show worse performance as they must return to the commander agent's position each time to update their beliefs.

\subsubsection{Commander (non-broadcast communication) Scenario}

This scenario shows that the Selective model's team performance is better than the baseline models. With comparable total number of actions and communications, it is shown a better performance in terms of completion time.

\subsubsection{Blocked Cells Scenario}

 This scenario shows that the No Communication model achieves a marginally better performance than other models. Agents tend not to communicate due to having common knowledge and a shared plan at the beginning of the task, in which an agent will not follow another agent to the same state.

\begin{table*}
\centering
   \caption{Experimental results of the Gridworld domain}     \label{tbl:1}
  \includegraphics[scale=0.8]{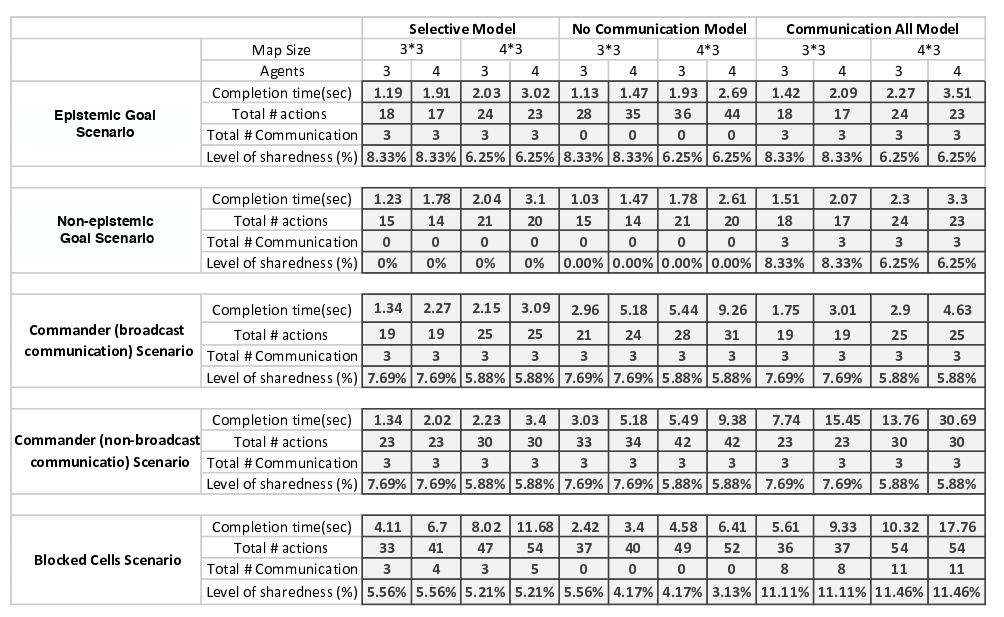}
\end{table*}

\begin{table*}
\centering
   \caption{Experimental results of the BW4T domain}     \label{tbl:2}
  \includegraphics[scale=0.8]{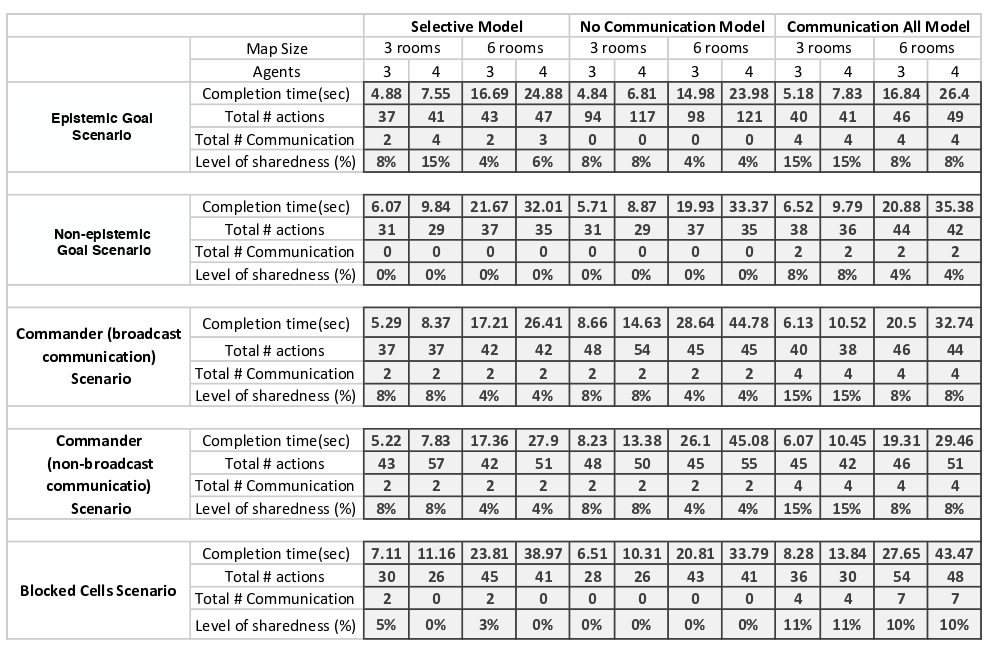}
\end{table*}
  
%%----------------------------------------------------					
\subsection{Discussion}

This study identified the effectiveness of the Selective communication model that contributed to team performance across different task scenarios compared to the No Communication and Communication All baseline models. In some task scenarios, the Selective model shows more sensible results for the BW4T domain than the Gridworld in terms of the total number of communications and level of sharedness. For instance, in the epistemic goal scenario, communication's selectivity is more apparent with different values across different team compositions and map sizes, unlike its results for the Gridworld domain. This may be because the task goal of the BW4T is to search and retrieve two specific color blocks from four blocks on the map. Thus, we determined that agents should communicate just what is relevant to the goal.

In addition, we found that epistemic planning helps more to solve problems that have goals in an epistemic form as it is the only way to solve such problems. However, it is also useful to solve problems whose goal is non-epistemic, such as blocked cell scenarios. Agents using the No Communication model achieved comparable performance in tasks having a non-epistemic goal. This could result from having common knowledge at the beginning of the task and a shared plan preventing an agent from following another agent to the same state. It is an important finding because it indicates that the Selective model may be particularly more beneficial for effective teamwork when presenting epistemic goals. One possible explanation for these results is that where there is no need to communicate (i.e., update other beliefs as a part of the goal), agents using the No Communication model will achieve comparable performance. This is because a team coordinates its actions implicitly based on the assumption that action effects are deterministic. \citeauthor{engesser2017cooperative} \cite{engesser2017cooperative} stated that relying on an epistemic planner's ability to plan for each agent as if it were another can elegantly solve problems without the necessity of explicit coordination. We can clearly see the importance of such a planning model in domains where there is a risk of information being explicitly communicated, for instance, in military operations. 

%----------------------------------------------------------------------------------------

\section{Conclusion and Future Work}

This paper aims to evaluate the efficiency of using a dynamic epistemic logic (DEL)-based planning approach to model communication processes through simulated experiments of multi-agent cooperative situations. The focus is mainly on search problems motivated by applications, such as search and rescue teams. Simulation domains were given to test the effects of the proposed model (Selective communication model) on team performance in incomplete information about the domain. In order for team members to work effectively towards achieving the tasks, they aligned the SMM that keeps track of consistent beliefs. They thus cooperated and made communication decisions effectively.

This paper modeled a communication process to be a part of planning as a natural action, in which its effect is to update agents' beliefs about the world. Thus, the planner communicates and keeps track of the SMM in one compact model. Our first step was to empirically evaluate how the communication model impacts team performance by characterizing the relationship between the degree of shared knowledge, communication actions, and team performance. We then compared our proposed model, the Selective model, with two baseline models generated from the proposed model. We found that the Selective model contributed more to team performance across all task scenarios than baseline models. This study is a first step towards the empirical evaluation of modeling communication explicitly as a part of planning in cooperative tasks and assessing its impact on team performance.

However, there are some limitations to our study that could be a further future area of work. First, we considered only two relatively-simple environments; looking at domains in which there are many competing goals and agents who must choose these goals would be a valuable extension. A further extensive evaluation is required for our proposed model using some simulation environments such as the Blocks World for Team (BW4T) simulator \cite{johnson2009joint}. We could also benchmark it against prior work of modeling communication proposed by \citeauthor{unhelkar2016contact} \cite{unhelkar2016contact}.

As we only ran one planner, we could get different results with different planners. Although the planner that we have used tends to minimize the cost, it was not optimal. Furthermore, as the epistemic planner used can reason nested beliefs (i.e., \ levels of knowledge) where it is bounded to three \cite{muise2015planning}, we only evaluated level one of knowledge, that is what agents believe about the world. Future work is needed to evaluate level two, and three of knowledge \cite{de2014theory}, and assess its impact on modeling communication and performing tasks.

We mainly focused on the communication content in terms of beliefs. A mental model consists of different components, such as intentions, goals, and beliefs, each with its weight and power to team performance \cite{jonker2011towards}. Results from prior work of measuring team performance with sharedness levels of various mental model components proved their contribution to team performance \cite{harbers2012measuring}. We could further investigate the impact of communicating different mental model components, such as intentions and goals, on team performance. 

Finally, we evaluated the Selective model on team performance based on the assumptions that agents are cooperative and commonly know the effects of actions. One avenue for further study would be to evaluate that model where agents are in a non-cooperative mode in which their goals are not shared.

%\section*{Disclosure Statement}
%We have no conflicts of interest to disclose. 

\smallskip

\bibliographystyle{IEEEtranN}
\bibliography{ap-article.bib}

%----------------------------- END BODY OF TEXT -----------------------------

%\end{multicols}

\end{document}